\documentclass[cmt]{svjour}

\usepackage{graphics}
\begin{document}

\title{Dynamics and Thermodynamics of a model with long-range interactions}
\author{Alessandro Pluchino, Vito Latora and Andrea Rapisarda 
}
\institute{Dipartimento di Fisica e Astronomia, Universit\'a di Catania,
and INFN Sezione di Catania,\\
Via S. Sofia 64,  I-95123 Catania, Italy}

\mail{A. Rapisarda (\email{andrea.rapisarda@ct.infn.it})}
 
\date{Received: July 8, 2003/ Accepted: October 27, 2003\\
Communicated by M. Sugiyama}

\maketitle
\begin{abstract}
The   dynamics and the thermodynamics of particles/spins
interacting via long-range forces display several unusual
features with respect to systems with short-range interactions.
The Hamiltonian Mean Field (HMF) model, a Hamiltonian system of N
classical inertial spins with infinite-range interactions 
represents a paradigmatic example of this class of systems.
The equilibrium properties of the model can be derived analytically in 
the canonical ensemble: in particular the model shows a second order 
phase transition from a ferromagnetic to a paramagnetic phase. 
Strong anomalies are observed in the process of relaxation 
towards equilibrium for a particular class of out-of-equilibrium 
initial conditions. In fact the numerical simulations show  
the presence of quasi-stationary state (QSS), i.e. metastable states which become 
stable if the thermodynamic limit is taken before the
infinite time limit. The QSS differ  strongly from 
 Boltzmann-Gibbs equilibrium states: they exhibit negative specific heat, 
vanishing Lyapunov exponents and weak mixing, non-Gaussian velocity 
distributions and anomalous diffusion, slowly-decaying correlations and aging.
Such a scenario provides strong hints for the 
possible application of Tsallis generalized thermostatistics.
The QSS have been recently interpreted as a spin-glass phase of the model. This link 
indicates another promising line of research, which is not alternative to the previous one.
\keywords{Phase transitions; Hamiltonian dynamics; Long-range interaction; Out-of-equilibrium
statistical mechanics}
\PACS{05.70.Fh,89.75.Fb,64.60.Fr,75.10.Nr }
\end{abstract}

\section{Introduction}
\label{intro}

Since the original paper by Ising \cite{ising}, magnetic models
on a lattice have been extensively used over the
years to investigate the statistical physics of
interacting many-body systems.
In particular, many generalizations of the Ising model have been
proposed. Among them, also models with long-range interactions.
The thermodynamics and the dynamics of systems of particles
interacting with long-range forces are particularly interesting
because display a series of anomalies with respect to systems
with short-range interactions\cite{leshouches} .
By long-range interaction it is usually intended that the modulus of
the potential energy decays, at large distance, not faster than
the inverse of the distance to the power of
the spatial dimension.
The main reason of the observed anomalies is that systems
with long-range forces in general violate {\it extensivity} and
{\it additivity}, two basic properties to derive the thermodynamics
of a system.
Extensivity means that the thermodynamic potentials scale
with the system size, i.e. that the specific thermodynamic potentials
(the thermodynamic potentials per particle) do not diverge in the
thermodynamic limit.
Additivity means that if we divide the system into two macroscopic
parts, the thermodynamic potentials of the whole
system are, in the thermodynamic limit, the sum of those of the
two components.
Though extensivity can be artificially restored by an ad-hoc
introduction of a N-dependent coupling constant (the so called Kac's
prescription \cite{kac}) in the interaction, the problem
of non-additivity is still present \cite{leshouches}.
Thus, the study of long-range magnetic systems on lattices
can give insights on the statistical properties
of long-range potentials, and is
therefore useful to investigate the statistical mechanics and
the dynamics of such systems.\footnote{Very often, as in this case, the term "nonextensivity" is used to refer to the "non-additivity" property.}

In this paper we study the $HMF - \alpha$ model, a model
of planar spins on a  d-dimensional lattice with couplings that
decay as the inverse of the distances between spins raised to
the power $\alpha$ \cite{alfaxy,giansanti0}.
When $\alpha$ is smaller than $d$ additivity does not
hold and the system shows a series of anomalies.
Even ensemble equivalence, whose proof is based on the possibility
of separating the energy of a subsystem from that of the whole,
might not be guaranteed in that limit.

In particular we focus on $\alpha=0$. In such a case the
model reduces to a mean field model\cite{antoni,lat2,lat3,lrt_pre,monte,plr1,moya,yama}, since a spin interacts equally
with all the others independently of their position on the lattice.
This case is extremely important since it has been proved that all the cases
with $\alpha /d \leq 1 $ \cite{alfaxy,giansanti0} can be reduced to it. 
 Moreover the dynamical behavior of the model with $\alpha=0$ is also a 
representative example of other nonextensive systems 
\cite{alfaxy,giansanti0,lrt_pre,cabral,lh02,lj,nobre}.
We will discuss the equilibrium phase and the dynamical anomalies found in an 
energy region before the second-order critical point when one studies the 
relaxation towards equilibrium. The connections with Tsallis 
generalized thermodynamics \cite{tsa1,lrt_pre,lh02,plr1,cho} 
and with glassy systems\cite{bou,plr_vetri} will be also addressed.  

The paper is organized as follows. We introduce the model in section 2,
the equilibrium thermodynamics is presented in section 3, while the anomalous
dynamical behavior and its possible theoretical interpretation is discussed in section 4. Conclusions and future prespectives
are presented   in section 5.

\section{The HMF model}
\label{sec:1}
The $HMF-\alpha$ model has been introduced in \cite{alfaxy}
and describes a system of classical bidimensional spins (XY spins) 
with mass $m=1$. The Hamiltonian is:
\begin{equation}
\label{hmfalpha}
 H_{\alpha} = K+V_{\alpha}
= \sum_{i=1}^{N} \frac{p_i^2}{2}  +  \frac{\epsilon}{2\tilde{N}}
 \sum_{i\neq j}^{N} \frac {1-\cos(\theta_i-\theta_j)}
 {r_{ij}^\alpha} \, .
\end{equation}
with $\epsilon = \pm 1$.
The $N$ spins are placed at the sites of a generic
$d$-dimensional lattice, and each one is represented by
the conjugate canonical pair $(p_i,\theta_i)$,
where the $p_i$'s are the momenta
and the $\theta_i$'s $\in[0,2\pi)$ are the angles of
rotation on a family of parallel planes, each one defined at each lattice
point. The interaction between rotators $i$ and $j$
decays as the inverse of their distance $r_{ij}$ to the
power $\alpha\geq 0$.

Such Hamiltonian is extensive if the thermodynamic limit (TL)
$N\rightarrow\infty$ of the canonical partition function $(\ln Z)/N$ exists
and is finite. This is assured for each $\alpha$ by the presence of the
rescaling factor $\tilde{N}$ in front of the double sum of the potential
energy. $\tilde{N}$ is a function of the lattice parameters $\alpha,d,N$
which is proportional to the range $S$ of the interaction defined by
\cite{alfaxy,giansanti0}:
\begin{equation}
\label{range}
\tilde{N}\propto S=\sum_{j\neq i} \frac{1}{r_{ij}^\alpha} \, .
\end{equation}
The sum is independent of the origin $i$ because of periodic conditions.
Then, for each $\alpha$, $V_{\alpha}$ is proportional to $N$. When $\alpha>d$, which
we call here the \textit{short-range} case, $S$ is finite in the TL
\cite{giansanti0}, and things go as if each rotator interacted with a finite
number of rotators, those within range $S$. On the contrary when
$\alpha<d$, which we consequently call the \textit{long-range} case, $S$
diverges in the TL and the factor $1/\tilde{N}$ in
(\ref{hmfalpha}) compensates for this.

For $\alpha=0$ the model (\ref{hmfalpha}) reduces to
the model introduced originally in Ref.\cite{antoni}
and called HMF model. The Hamiltonian of the HMF model is:
\begin{equation}
H_0 = K + V_0 = \sum_{i=1}^N \frac{p_i^2}{2} + \frac{\varepsilon}{2N}
\sum_{i,j=1}^N [1-\cos(\theta_i - \theta_j)]~~~~.
\label{model0}
\end{equation}
\noindent
 This model can be seen as classical $XY$-spins with infinite
range couplings, or also as representing particles moving 
on the unit circle. In the latter interpretation
the coordinate $\theta_i$ of particle $i$ is its position on the circle 
and $p_i$ its conjugate momentum. 
For $\varepsilon>0$, particles attract each other
or, equivalently speaking, spins tend to align (ferromagnetic case), while for
$\varepsilon<0$, particles repel each other and spins tend to
anti-align (antiferromagnetic case) \cite{leshouches}. 
At short distances, we can
either think that particles cross each other or that they collide
elastically since they have the same mass.
For simplicity of the notation in the following we omit the subscript $\alpha=0$.
One
can introduce the mean field vector
\begin{equation}
{\bf M} = M {\rm e}^{i \phi} = \frac{1}{N} \sum_{i=1}^N {\bf m}_i
\label{m0}
\end{equation}
where ${\bf m}_i =(\cos \theta_i, \sin \theta_i)$. Here, $M$ and $\phi$
represent the modulus and the phase of the order parameter, which
specifies the degree of clustering in the particle interpretation,
while it is the {\em magnetization} for the $XY$ spins. Employing
this quantity, the potential energy can be rewritten as a sum of
single particle potentials $v_i$
\begin{equation}
V = \frac{1}{2} \sum_{i=1}^N v_i  \qquad {\rm with} \qquad v_i =
1- M \cos(\theta_i -\phi) \quad . \label{v0}
\end{equation}
It should be noticed that the motion of each particle is coupled to all
the others, since the mean-field
variables $M$ and $\phi$ are determined at each time $t$ by the
instantaneous positions of all particles.

\section{Thermodynamics}
\label{sec:2}
In this section we review the equilibrium solution of the HMF model   
derived in the canonical ensemble in ref. \cite{antoni}. 
The equilibrium solution of general case $HMF - \alpha$ model 
can be found in refs. \cite{giansanti0}.
In the canonical ensemble we need to evaluate the partition function: 
\index{canonical!partition function}
\begin{equation}
Z = \int d^N p_i d^N \theta_i \exp{(-\beta H)}~~~~,
\label{z0}
\end{equation}
where $\beta = 1/(k_B T)$, $k_B$ is the Boltzmann constant and
$T$ is the temperature. The integration domain is extended
to the whole phase space. 
It is suitable to factorize the partition function in a 
kinetic part
\begin{equation}
Z_K = \int_{-\infty}^{\infty} d^N p_i \exp{\left(-\frac{\beta}{2} \sum_i p_i^2\right)} = \left( \frac{2\pi}{\beta} \right)^{N/2} ~~~,
\end{equation}
and a potential one
\begin{equation}
Z_V = \exp{\left[\frac{-\beta \varepsilon N}{2} \right]} \int_{-\pi}^\pi d^N
\theta_i \exp{\left[\frac{-\beta \varepsilon}{2N} \sum_{i,j} cos(\theta_i - \theta_j)\right]}
~~~~.
\end{equation}
We shall consider only the ferromagnetic condition, i.e. $\varepsilon = 1$. We have 
\[
\sum_{i,j} cos(\theta_i - \theta_j) 
= \left(\sum_{i} cos\theta_i \right)^2 + \left(\sum_{i} sin\theta_i \right)^2 = \left| \sum_{i} \bf m_i \right|^2~~~~,
\]
thus eq. (\ref{z0}) can be rewritten as
\begin{equation}
Z = C \int_{-\pi}^\pi d^N \theta_i \exp{\left[\frac{-\beta N}{2} {\bf M}^2 \right]}~~,
\label{zz0}
\end{equation}
where
\begin{equation}
C = \left( \frac{2\pi}{\beta} \right)^{N/2} \exp{\left[\frac{-\beta N}{2}\right]}~~.
\end{equation}



\begin{figure}[htbp]
\begin{center}
\resizebox{0.85\textwidth}{!}{
  \includegraphics{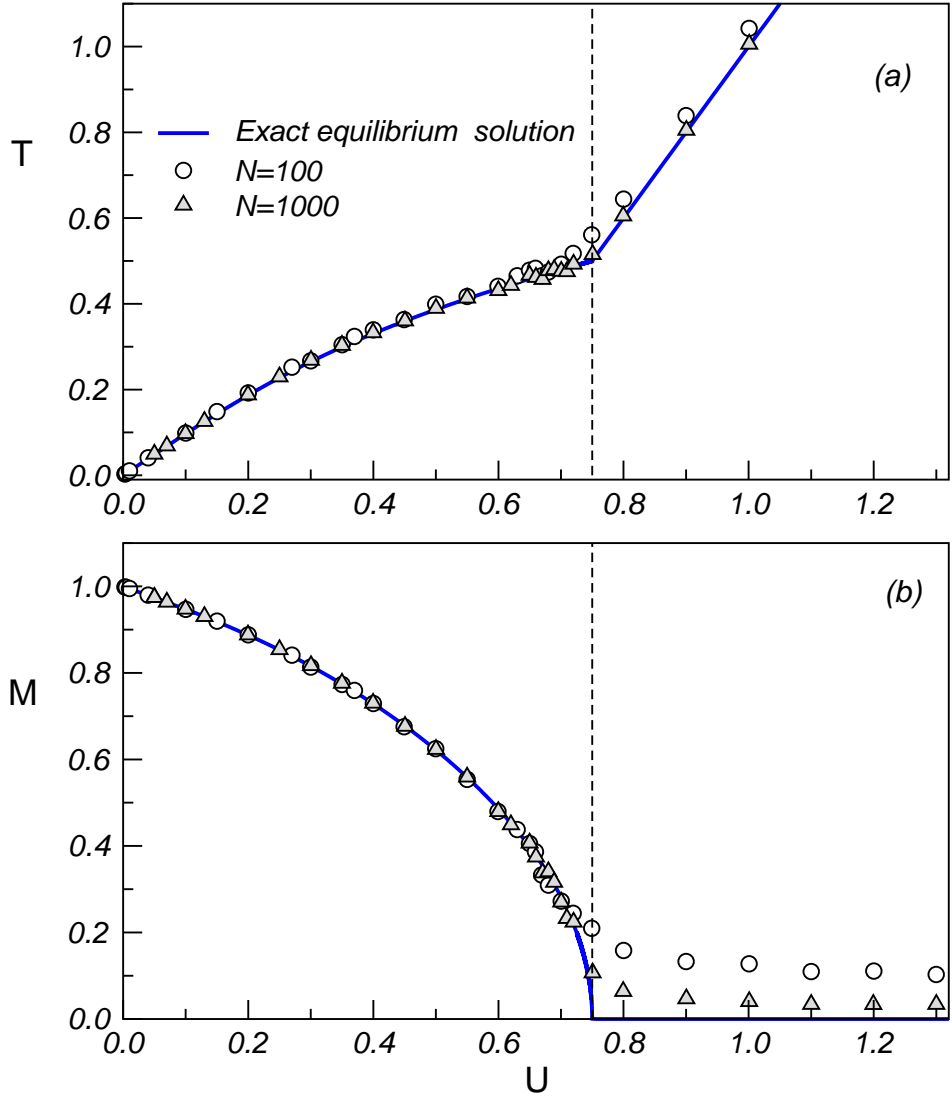}
}
\caption{
Temperature $T$ and magnetization $M$ as a function of the
energy per particle $U$ in the
ferromagnetic case.
Symbols refer to equilibrium molecular dynamic simulations
 for $N=10^2$ and $10^3$, while the solid
lines refer to the canonical equilibrium prediction obtained analytically, see text. The vertical dashed line indicates the
critical energy density  located at $U_c=0.75$ and 
$\beta_c= \frac{1}{T_c}=2$.}
\label{fig:1}
\end{center}
\end{figure}


In order to evaluate this integral, we use the
Gaussian identity
\begin{equation}
\exp{\left[ \frac{\mu}{2} \bf x^2\right]}=
\frac{1}{\pi} \int_{-\infty}^\infty
d{\bf y} \exp{[-{\bf y}^2+\sqrt{2 \mu} ~{\bf x}
\cdot {\bf y}]}~~~,
\label{gau}
\end{equation}
where ${\bf x}$ and ${\bf y}$ are two-dimensional vectors and
$\mu$ is positive. We can therefore rewrite Eq.~(\ref{zz0}) as
\begin{equation}
Z = \frac{C}{\pi}
\int_{-\pi}^\pi d^N \theta_i \int_{-\infty}^\infty
d{\bf y} \exp{[-{\bf y}^2+\sqrt{2 \mu} {\bf M}
\cdot {\bf y}]}
\label{j0}
\end{equation}
and $\mu = \beta N$. We use now definition~(\ref{m0}) and
exchange the order of the integrals in (\ref{j0}), factorizing the
integration over the coordinates of the particles. Introducing
the rescaled variable $ {\bf y} \to {\bf y} \sqrt{N/2\beta}$,
one ends up with the following expression for $Z$:
\begin{equation}
Z = \frac{N C}{2 \pi \beta}
\int_{-\infty}^\infty d{\bf y} \exp{\left[-N \left( \frac{y^2}{2
\beta} -\ln\left(2\pi I_0(y)\right) \right) \right] }~~~,
\label{j1}
\end{equation}
where $I_0$ is the modified Bessel function of order $0$ and $y$
is the modulus of ${\bf y}$. Finally, integral~(\ref{j1}) can be
evaluated by employing the saddle point technique in the
thermodynamic limit, i.e. for $N \to \infty$.
In this limit, the
Helmholtz free energy
 per particle $f$ reads as:
\begin{equation}
\beta f=-\lim_{N \to \infty} \frac{\ln Z}{N}=
-\frac{1}{2}\ln\left(\frac{2 \pi}{\beta}\right) +\frac{
\beta }{2} +\max_y\left[\frac{y^2}{2\beta}-\ln(2\pi
I_0(y))\right]\quad ~~~,
\label{freehmf}
\end{equation}
while the maximum condition leads to the following consistency equation:
\begin{equation}
\frac{y}{\beta}=\frac{I_1(y)}{I_0(y)}\quad ~~~~,
\label{cons}
\end{equation}
where $I_1$ is the modified Bessel function of order $1$.
Eq.(\ref{cons}) is the analogous, in the XY model, of the Curie-Weiss equation obtained by solving
the Ising model in the mean field approximation.
For $\beta \leq \beta_c = 2$ it presents only the solution $\bar y=0$, that is unstable . At $\beta=\beta_c$, i.e. below the critical temperature $T_c=0.5$ ($k_B=1$), two new stable symmetric
solutions appear through a pitchfork bifurcation and a discontinuity
in the second derivative of the free energy is present, indicating a
second order phase transition\index{phase!transition}.
These results are confirmed by an analysis of the order
parameter \footnote{This is obtained by adding to the Hamiltonian
an external field and taking the derivative of the free energy
with respect to this field, evaluated at zero field.}
\begin{equation}
M= \frac{I_1(\bar y)}{I_0(\bar y)} \quad. \label{mag0}
\end{equation}
The magnetization $M$ vanishes continuously
at $\beta_c$ (see Fig.~\ref{fig:1}(b)). Thus, since $M$
measures the degree of clustering of the particles, we have
a transition from a clustered phase when
$\beta>\beta_c$ to a homogeneous phase when $\beta<\beta_c$.
The exponent which characterize the behavior of the magnetization
close to the critical point is $1\over2$ as expected for a mean field
model \cite{antoni}.
One
can obtain also the energy per particle vs temperature and magnetization
\begin{equation}
U= \frac{\partial(\beta f)}{\partial \beta}=
\frac{1}{2\beta}+\frac{1}{2}\left(1-M^2\right)~~,
\label{u0}
\end{equation}
the so called  {\it caloric curve}, which is reported in Fig. \ref{fig:1}(a).
\begin{figure}[htbp]
\begin{center}
\resizebox{0.85\textwidth}{!}{
  \includegraphics{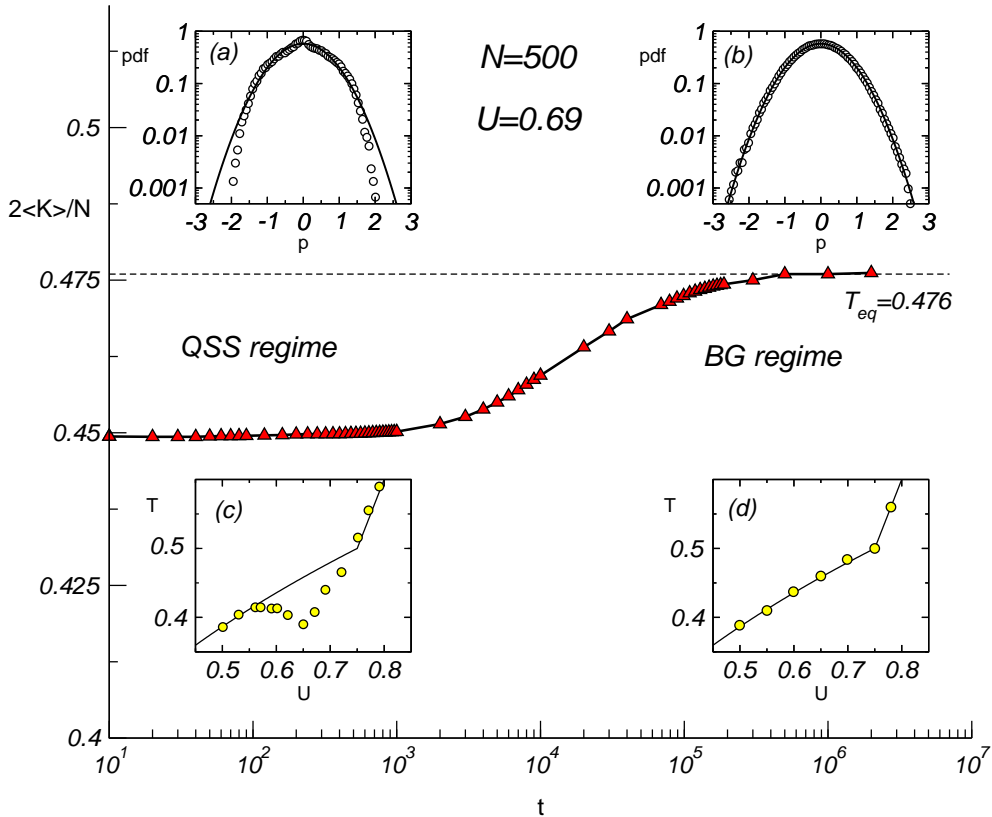}
}
\caption{ Microcanonical numerical simulations
for $N=500$ and energy density $U=0.69$.  In the main part of the
figure we plot
 twice the  average kinetic energy per particle (which gives the temperature) as a function of time (filled triangles). We can easily
 distinguish  a long matastable plateau (QSS regime)
   preceeding  the relaxation towards the Boltzmann-Gibbs
 equilibrium temperature (BG regime). In the  BG regime, one
 finds as expected, a very good agreement with the equilibrium thermodynamics
 value for the temperature, panel (d). In this regime the
  velocity pdfs reported in panel (b) are Gaussians.
 At variance, in the QSS region, we observe strong deviations from the  expected equilibrium   temperature. Here the specific heat becomes negative, panel(c), and the
 velocity pdfs, reported in panel (a),
   are very different from  the Gaussian equilibrium curve, reported as a full line for comparison. See text for further details.}
\label{fig:2}
\end{center}
\end{figure}
\\
In  Fig.1 we report temperature and magnetization vs the energy density. 
The full curves are the exact results obtained in the canonical 
ensemble, the Boltzmann-Gibbs (BG) equilibrium. 
 The numerical simulations (symbols) were performed at fixed total energy
({\em microcanonical} molecular dynamics)
by starting the system close to the equilibrium, i.e. with an initial Gaussian 
distribution of angles and velocities, and reproduce the theoretical curves. 
This is already true for small system sizes as $N=100$,    
apart from some finite size effects in the homogeneous
phase \cite{leshouches}. 
As we shall see in the next section the scenario is very different 
when the system is started with strong out-of-equilibrium initial conditions: 
in such a case the dynamics does have difficulties in  reaching
 the BG equilibrium  and shows a series of anomalies.

\section{Dynamics}
\label{sec:3}
The dynamics of each particle obeys the following
pendulum equation of motion:
\begin{equation}
\ddot \theta_i= \sum_{j=1}^{N} sin(\theta_j-\theta_i)
= - M \sin(\theta_i-\phi) \quad , ~~~~ i=1,.....,N
\label{pend}
\end{equation}
where $M$ and $\phi$ have a non trivial time dependence, related
to the motion of all the other particles in the system.
Equations~(\ref{pend}) can be integrated numerically, 
see for example refs. \cite{antoni,lat2} for technical details. 
 In this  section we show that, in an energy range from $U=0.5$ up to 
$U_c=0.75$, when the system is started with strong 
out-of-equilibrium initial 
conditions, the model has a non trivial dynamical 
relaxation to the Boltzmann-Gibbs (BG) equilibrium. 
The class of out-of-equilibrium initial conditions we consider, 
called {\it water bag} initial conditions, consists in $\theta_i =0 ~\forall i$ 
and the momenta uniformly distributed (according to the total energy  density $U$). 
In Fig.2 we report, for U=0.69 and N=500, the time evolution 
of $2<K>/N$ (where $<K>$ denotes the time averaged kinetic energy), a quantity that
coincides with the temperature T.
As expected, the system does not relax immediately to the BG equilibrium, 
but rapidly reaches a quasi-stationary state (QSS) corresponding to
a temperature plateau situated below the canonical prediction (dotted curve).
The T vs U plot (caloric curve), shown in inset (c) for the QSS regime, confirms 
a large disagreement with the equilibrium prediction (inset d). 
Furthermore, it turns out that the system remains trapped in such a state for a time
that diverges with the size $N$ of the system \cite{lrt_pre}.
This means that, if the thermodynamic limit is performed before the
infinite-time limit, the QSS become stable and the system never relax to the
BG equilibrium, exhibiting different equilibrium properties characterized
by non-Gaussian velocity distributions (see inset a) \cite{lrt_pre}. 
Such velocity distributions have been fitted in 
Ref.\cite{lrt_pre} and have been shown to be in agreement with 
the prediction of the Tsallis' generalized thermodynamics 
\cite{tsa1}. The latter  is a theoretical formalism well suited to describe 
all those situations where long-range correlations, weak mixing  and 
fractal structures in phase space are present \cite{cho}, such as for example
excited plasmas \cite{bog}, 
turbulent fluids \cite{beck,bc}, 
maps at the edge of chaos \cite{lyra,logvito,fulvio,stand}, 
high energy nuclear reactions \cite{wilk} and cosmic rays fluxes\cite{astro}.  
\\
The characteristics of the QSS have been studied in several papers:  
vanishing Lyapunov spectrum \cite{lrt_pre,cabral,lh02},  
negative microcanonical specific heat\cite{lat3},  
dynamical correlations in phase-space \cite{lrt_pre,plr1},  
L\'evy walks and anomalous diffusion\cite{lat2}.
Recently also the validity of the zeroth principle of 
thermodynamics  has been numerically demonstrated for these metastable states\cite{moya}.
\\
In the following of this paper we focus on the 
study of anomalous diffusion, long-range correlations and the 
spin-glass phase.

\subsection{Anomalous Diffusion} 
The link between relaxation to the BG equilibrium, 
and anomalous diffusion was studied in Ref.\cite{lat2}.
%
\begin{figure} 
\begin{center}
\resizebox{0.85\textwidth}{!}{
 \includegraphics{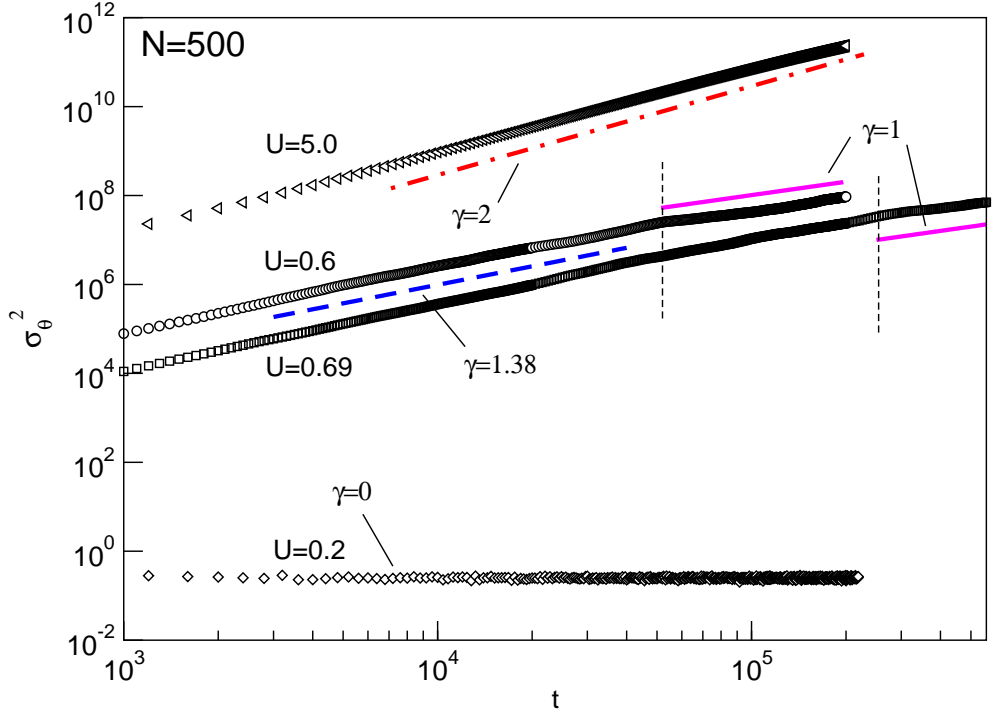}
}
\caption{
We plot the time evolution of the variance for the angular displacement, see
eq.(18), for
$N=500$ and various energy densities. A ballistic diffusion behavior, $\gamma=2$
is found for overcritical energies. No diffusion, $\gamma=0$,
is of course observed at very small energies, $U<0.25$. Anomalous diffusion
$\gamma\approx 1.4$
is obtained   for the case $U=0.6$ and $U=0.69$. This behavior is seen
 only for a time interval equal to the duration of the QSS regime, indicated by vertical dashed lines (see also the previous figure). When relaxation to equilibrium is attained
  the diffusion becomes again normal, $\gamma=1$. This relaxation time depends on $U$ and diverges  linearly with $N$ \cite{lrt_pre}.}
\label{fig:3}
\end{center}
\end{figure}
%
%
In that paper the authors studied the variance for the angular displacement 
defined as
\begin{equation}
\sigma_{\theta}^2(t) = \frac{1}{N} \sum_{i=1}^N [\theta_i(t)-\theta_i(0)]^2~~~,
\label{MSD}
\end{equation}
which tipically depends on time  as $\sigma^2 (t)\propto{t^\gamma}$. 
The diffusion is anomalous when $\gamma \neq1$ and, in particular, 
it is called {\it subdiffusion} if $0<\gamma<1$ and {\it superdiffusion} if $1<\gamma<2$.
In ref.\cite{lat2} the authors found the existence of an anomalous 
superdiffusive behavior below the critical energy, 
connected to the presence of the QSS. 
Superdiffusion turns into normal diffusion after a crossover time 
$\tau_{crossover}$ that coincides with the time $\tau_{relax}$ needed for the QSS 
to relax to the BG  equilibrium. We illustrate  this behavior in Fig.\ref{fig:3}.
In particular we show for the case $N=500$ and various energy densities the
time evolution for  $\sigma_{\theta}^2(t)$. No diffusion, $\gamma=0$,
is observed for very low energy, while one gets ballistic motion in the  overcritical
energy region, $\gamma=2$. On the other hand, superdiffusion is found in the energy
region $U=0.5-0.75$ with $\gamma\approx 1.4$. We plot the cases $U=0.6$ and $U=0.69$. As expected, however,  diffusion becomes
again normal, $\gamma=1$, after complete relaxation. In the figure we indicate with vertical dashed lines these
relaxation times $\tau_{relax}$ for $U=0.6$ and $U=0.69$.
 Lines with different slopes are also reported to indicate the  different diffusion regimes.

Notice that diffusion starts around $U=0.3$ when particles start to be 
evaporated from the main cluster \cite{antoni}.
 
This result can be interpreted in the following way. When started with water-bag 
initial condition ($M=1$) the system immediately decays into the QSS plateau with
$M \approx 0$. In Fig.2 this initial part is not shown. Then a very slow microscopic 
relaxation occurs since the force exerted on each single spin/particle is almost zero \cite{lrt_pre}. 
 During the slow relaxation process, many small rotating clusters  
on the unitary circle are continuosly formed and compete each other by 
trapping the particles in order to reach a 
configuration compatible with the final BG equilibrium state. 
Thus, particles remains trapped for a while in the clusters, until they finally
succeed to escape again and so on.  
Trapping times and escape times obey power-law decays with characteristic exponents
that allow to relate anomalous diffusion with L\'evy walks.
See ref. \cite{lat2} for more details.
Such a mechanism of trappings and L\'evy walks disappears when the system finally
reaches equilibrium since at that time  only one large cluster is present (the system is below the critical
point in the ferromagnetic phase).  

Anomalous diffusion can be obtained within a generalized Fokker-Planck equation
which generates Tsallis distributions with an entropic index $q$, see refs.\cite{tsadif,beck3}. One
 can extract the following relationship between the exponent $\gamma$ which characterizes anomalous diffusion and the entropic index $q$ of Tsallis thermostatistics
\begin{equation}
\gamma=\frac{2}{3-q}~~~.
\end{equation}
Considering the value of $\gamma \approx 1.38-1.4$ observed in our case, we would expect then a value of the entropic index  $q \approx 1.55-1.58$ which should characterize the dynamical anomalies of 
the HMF model. Actually, this is not the value found in ref.\cite{lrt_pre} for the velocity pdfs, where
we obtained only an effective entropic index. However this value
  is in good agreement
 to what has been 
found for the decay of the velocity correlation functions which will be examined in the next section.

\subsection{Slow relaxation and aging} 

The global effects of the competition between magnetic clusters discussed in the 
previous subsection prevents the system from exploring all the available phase space. 
Thus one observes  a sort of {\it dynamical frustration} which  suggests also
an interesting connection with the {\it weak-ergodicity breaking}
scenario typical of glassy systems. Such a scenario, introduced by Bouchaud et al.
\cite{bou}, generally occurs when the phase-space of the system
we consider is not {\it a-priori} broken into mutually inaccessible regions, but
the system remains confined only into a restricted part of it: consequentely  
one finds slowly decaying correlation functions and aging, i.e.  
the presence of strong memory effects that depend on the history of the system.
This is just what happens in the HMF model.  
\\
The velocity autocorrelation functions in the QSS regime have been studied 
in ref\cite{plr1}. 
In its simplest form the autocorrelation function of the particles velocity 
can be written as \cite{yama}
\begin{equation}
C_p (t) = \frac{1}{N} \sum_{i=1}^N p_j (t) p_j (0)~~~. 
\end{equation}
If we want to take into account several dynamical realizations (events)
in order to obtain better averaged quantities, it is possible to use the 
following alternative definition of the autocorrelation function 
\begin{equation}
{ C_p}(t)= {< {{{\bf P}(t) \cdot {\bf P}(0)}}> -
{{<{\bf P}(t)>\cdot <{\bf P}(0)>}} \over {\sigma_p(t)\sigma_p (0)}}~~~,
\label{C_p}
\end{equation}
where ${\bf P}=(p_1,p_2,...p_N)$ is  the   N-component  velocity vector
and the brackets $<...>$ indicate the
average over different events, while  $\sigma_p(t)$ and $\sigma_p(0)$
are the standard deviations at time $t$ and at the initial time.
Such a function has been numerically evaluated \cite{plr1} in the QSS region. 
The results are shown in Fig.4 for the case
$U=0.69$ and $N=1000$: a power-law tail is observed, this being  
a signature of long-range correlations.  
%
\begin{figure} 
\begin{center}
\resizebox{0.85\textwidth}{!}{
  \includegraphics{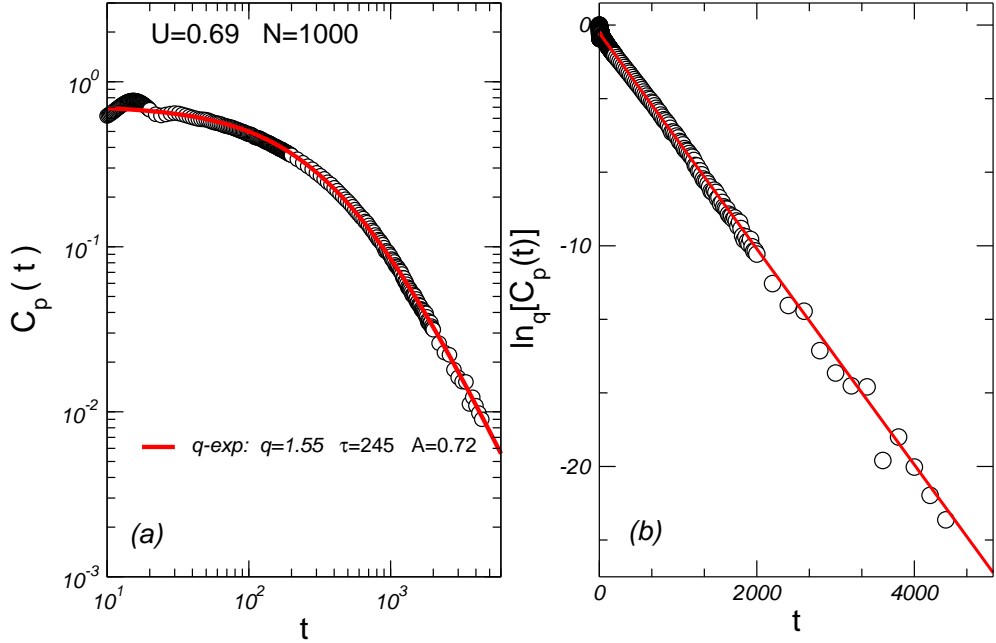}
}
\caption{
(a) We plot the  velocity correlation function  $C_p(t)$ vs. time t for the case $U=0.69$ and $N=500$ in the QSS region, open symbols.
The curve plotted as a full line is a q-exponential fit, with $q=1.55$.
(b) We plot here the q-logarithm of the curves and points
reported in (a), see text for further details.
}
\label{fig:4}
\end{center}
\end{figure}
%
\begin{figure}[htbp]
\begin{center}
\resizebox{0.85\textwidth}{!}{
  \includegraphics{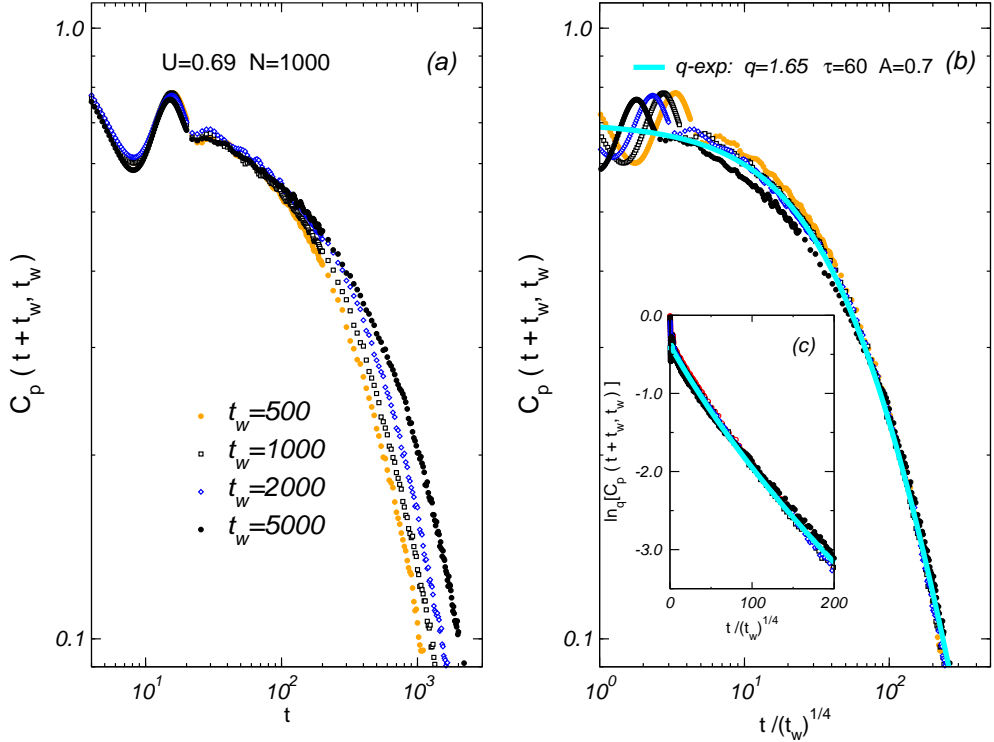}
}
\caption{
(a) For $U=0.69$ and $N=1000$, we plot the  two-times velocity autocorrelation function defined by eq.(25). Different delay times
$t_w$ are considered, open symbols. Aging, i.e. a marked dependence of the correlation function on $t_w$
is clearly evident. (b) The curves collape onto a  single one if an opportune scaling is performed, see text.
This behavior can be reproduced by a q-exponential fit also reported. We plot also  in (c) the q-logarithm
of the data drawn in (b).
}
\label{fig:5}
\end{center}
\end{figure}
%
The power-law tail and the initial 
saturation can be fitted by means of the  q-exponential relaxation function of 
the Tsallis' generalized thermodynamics \cite{tsa1}:
\begin{equation}
e_q(z)=  {\left[  1+(1-q) {z} \right]} ^{1\over(1-q)}~~~~,
\end{equation}
with $z={-y\over \tau}$. Here , 
$\tau$ is a characteristic time and $q$ is the entropic exponent
which takes into account the specific dynamical anomalies of the model into
exam. We multiply the curve by a constant renormalization factor A (the saturation value). In our case we get for $U=0.69$ and $N=1000$ a value $q=1.55$ with 
a saturation value $A=0.72$ and a characteristic time $\tau=245$.
By plotting the inverse function of the q-exponential, i.e. the q-logarithm defined
as 
\begin{equation}
ln_q(z)=  {{z^{1-q} -1 } \over {1-q}} ~~~,
\end{equation}
one can check the quality of the agreement between the theoretical expected
behavior and the numerical simulations. This is done in Fig.4(b).

To study the aging phenomenon we need to evaluate a two points correlation function: 
\begin{equation}
{ C_p}(t+t_w,t_w)= {<{{{\bf P}(t+t_w)\cdot {\bf P}(t_w)}}> -
{{<{\bf P}(t+t_w)> \cdot <{\bf P}(t_w)>}} \over {\sigma_p(t+t_w) \sigma_p(t_w)}}.
\end{equation}
where $t_w$ is the waiting time. This function has been evaluated in 
ref.\cite{monte,plr1}. In the QSS region,
one finds a strong dependence on the waiting time $t_w$
see Fig.5 (a) for the case $U=0.69$ and $N=1000$.
The  curves obey a precise scaling as that one observed for glasses.
Rescaling the time of the various correlation curves by
$t/t_w^\beta$ with $\beta=\frac{1}{4}$, we get a unique curve with
a very interesting power-law tail, see Fig.5 (b).
Also in this case as in the previous figure,
 one can reproduce nicely the time evolution  of the rescaled correlation
function with a
q-exponential curve. The best fit, reported in the figure, was obtained  for 
case $q=1.65$, $\tau=60$ and $A=0.7$. The q-logarithm is reported in panel (c)
in order to compare the quality of the fit.  Notice that this value of $q$
is not very different from the one extracted from the decay of  the 
autocorrelation function (22) and from that one extracted from the anomalous
diffusion. A more detailed investigation in order to deduce analitically 
the predicted value of the entropic index from the observed anomalies is 
in progress.

\subsection{Spin-glass phase}

In the long-range spin-glass models the aging phenomenon below a transition temperature is associated with the complex energy landscape characteristic 
of the frustrated models. 
The latter is  rich of metastable states
that play the role of dynamical traps which can confine
the system for a long time and cause the observed history-dependent slow relaxation 
dynamics\cite{bou,kurchan}.
The parallel with spin-glass systems can be made stronger by the 
introduction of a new order parameter for the QSS plateau, 
the $\it polarization~~ p$ \cite{plr_vetri}. The polarization  
represents the temporal average (over a time interval $\tau $ inside the QSS plateau) of the spin vectors, whose modulus is also 
averaged over the $N$ spins, i.e.
\begin{equation}
\label{pol}
{\it p}={1\over{N}} \sum_{i=1}^N  |<\stackrel{\vector(1,0){8}}{s_i}>_{\tau}|~~~~.
\end{equation}
Such an order parameter allows to characterize in a quantitative way the
dynamical freezing and frustration of the QSS regime by interpreting the latter 
as a spin-glass phase. In fact, in analogy
with the spin-glass phase of the long-range 
Sherrington-Kirkpatrick (SK) model \cite{sk}, 
 we have  observed that, in the QSS regime and  in the thermodynamic limit,
the magnetization vanishes as $M^{-\frac{1}{6}}$, but the polarization $p$
remains constant around a value equal to $\approx 0.24$, see Fig.2 of ref. 
\cite{plr_vetri}. Therefore $p$  allows  to distinguish between 
the disorderd QSS glassy phase and the high temperature one.

This result is very  interesting because, at variance with standard glassy systems, neither disorder nor frustration
are present {\it a-priori } in the elementary interactions of the HMF model. On the contrary,
 they emerge naturally as dynamical features of the model in QSS  region  and, like the other features seen before,
disappear when the system reaches finally the BG equilibrium.
The introduction of  the polarization order parameter, which connects the QSS dynamical frustration with the Edwards-Anderson order parameter of the SK model, opens a new 
perspective  to understand the true nature of metastability in long-range Hamiltonian systems in terms of an emerging glassy  behavior.

\section{Conclusions}
\label{sec:5}
In this paper we have briefly reviewed the dynamics and thermodynamics 
of the Hamiltonian Mean Field
model. This is a model that allows
to study long-range interactions in many-body Hamiltonian systems. 
The equilibrium properties of the HMF model can be derived analitically, though 
the model dynamics  presents very interesting anomalies
like metastability, superdiffusion, 
non-Gaussian velocity pdfs, long-range correlations, 
vanishing Lyapunov exponents and weak-ergodicity breaking 
in an energy  region below the critical point.
These anomalies are common to many  long-range systems. 
Recently it has been found that such anomalous regime can be 
characterized as a glassy phase. Therefore  
the model  represents also  a very interesting bridge between nonextensive systems
and glassy systems and in this respect it will deserve more detailed 
investigations in the future. The link
with the nonextensive thermostatistics formalism proposed by Constantino Tsallis 
is also a very promising and intriguing line of research.
All the anomalies point in that direction and the future
years will be crucial in order to establish this connection in a firm and rigorous way 
as it has recently been done for example in  unimodal maps.


\end{document}